\documentclass[11pt]{article}
\input{psfig}
\title{{\bf Galaxies and Genes: Towards an Automatic Modeling of 
  Interacting Galaxies}}
\author{Christian Theis, Christoph Gerds, \& Christian Spinneker\\
\vspace{0.1cm}\\
\normalsize Inst.\ f.\ Theoretische Physik und Astrophysik der Univ.\ Kiel,
  Olshausenstr.\ 40, 24098 Kiel, Germany
}
\date{}
\begin{document}
\maketitle
\def\bull{\vrule height .9ex width .8ex depth -.1ex}
\makeatletter
\def\ps@plain{\let\@mkboth\gobbletwo
\def\@oddhead{}\def\@oddfoot{\hfil\tiny
``Dwarf Galaxies and their Environment'';
International Conference in Bad Honnef, Germany, 23-27 January 2001}%
\def\@evenhead{}\let\@evenfoot\@oddfoot}
\makeatother

\begin{abstract}\noindent
    The main problems in modeling interacting galaxies are the extended
parameter space and the fairly high CPU costs of self-consistent
N-body simulations. Therefore, traditional modeling techniques suffer 
from either extreme CPU demands or trapping in local optima (or both). 
A very promising alternative approach are evolutionary algorithms which 
mimic natural adaptation in order to optimize the numerical models.
One main advantage is their very weak dependence on starting points
which makes them much less prone to trapping in local optima.
We present a Genetic Algorithm (GA) coupled with a fast (but not
self-consistent) restricted N-body solver. This combination
allows us to identify interesting regions of parameter space within
only a few CPU hours on a standard PC or a few CPU minutes on
a parallel computer. Especially, we demonstrate here the ability of 
GA-based fitting procedures to analyse observational data automatically,
provided the data are sufficiently accurate.
\end{abstract}
%
%
\section{Introduction}

   Interacting galaxies are among the most fascinating astronomical
objects in the universe. The fate of these galaxies span a wide range
from single distant encounters 
to close encounters which might end in a single merged system.
Typical morphologies include e.g.\ bridges between the interaction 
partners or tidal tails. These structures are usually the sites of
strong star formation resulting in dense star clusters or even
dwarf galaxy sized objects. Additionally, a central star burst might be 
triggered by tidally induced bars which funnel matter to the
galactic centre. This coupling between galactic dynamics and 
''microphysical'' processes (like star formation) provides a unique
tool for a deeper understanding of galactic evolution. 

   An important prerequisite for the analysis of dynamically induced
mechanisms is the
knowledge of the dynamics of the interaction itself. E.g.\ the age of
a tidally formed stellar system derived from a colour-magnitude diagram
should be related to the time passed by since the last
perigalactic passage(s) (though they
need not to be identical). The main difficulty for a detailed modeling
is the extended parameter space which contains orbital and structural
parameters of both interacting galaxies. Already a simple interaction
of a galactic disk with a galaxy described by a point mass needs 7
parameters. Covering this parameter space by a ''complete'' grid 
of 10 grid points per dimension requires more than 
4 years of CPU time on a GRAPE5 special purpose computer. More efficient
search strategies like gradient methods suffer from their dependence on
initial conditions and, thus, a possible trapping in local optima. 

  An alternative are {\it genetic algorithms} (GAs) 
(Holland 1975, Charbonneau 1995) coupled with fast N-body methods. Because
GAs need about $10^4$ or more simulations, a single simulation
should not exceed 10 CPU secs. Contrary to self-consistent 
simulations, {\it restricted N-body} simulations (Toomre \& Toomre 1972) 
can be performed in one or a few CPU seconds on a modern PC.
Hence, they allow for a fast investigation of
the parameter space including both, finding a good fit and testing
preferred models on their uniqueness (Wahde 1998, Theis 1999).

  So far, the GA-based analyses have been used to model artificial data
generated by restricted N-body simulations. By this, the uniqueness
of preferred interaction scenarios for NGC 4449 (Theis \& Kohle 2001) and 
NGC 4631 (Theis \& Harfst 2001) have been tested. 
In this paper, we investigate if and how observed data can be modeled 
{\it automatically} without the need of a known reference model
calculated by a restricted N-body simulation.
In the next section we describe the basics of genetic algorithms. 
In Sect.\ \ref{automaticfitting} we study two cases: First, we generate
an artificial FITS image from a self-consistent N-body simulation and
try to fit it by our GA. In the second test, we use HI observations 
as direct input (data from M 51). The results are summarized in 
Sect.\ \ref{summary}.
%
%
\section{Genetic Algorithm}
\label{geneticalgorithm}

\begin{figure*}[htbp]
  \centerline{
  \psfig{figure=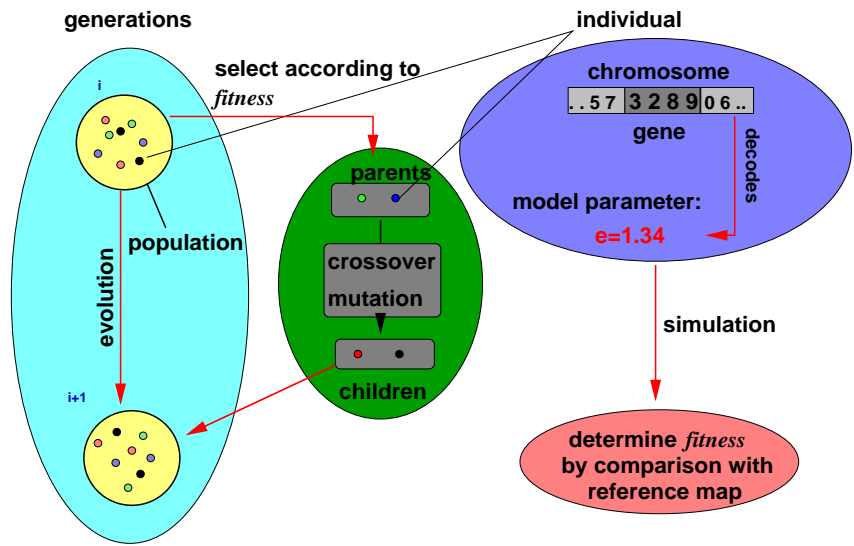,width=12.cm,angle=270.0}
  }
  \caption[]{Schematic diagram of the genetic algorithm approach}
  \label{fig_scheme}
\end{figure*}

The main idea of genetic algorithms is the application of evolutionary 
mechanisms in order to 'breed' a more and more adapted population
(Fig.\ \ref{fig_scheme}).
Each member of a population represents a single point in parameter
space, i.e.\ a N-body simulation with a given set of parameters.
The members are characterized by their fitness which quantifies the 
correspondance between the simulations and the reference map (observation or
numerical model). In order to determine the {\it 'parents'} 
two individuals are selected according to their fitness.
These parents are two points in parameter space.
The parameters of each individual are converted to a ''universal'' alphabet 
(here 4-digit numbers) and then combined to a single string, the
{\it 'chromosome'}. This chromosome is subject to a 
{\it cross-over}\footnote{Cross-over e.g.\ is realized by swapping the
ends of two chromosomes at a randomly chosen cross-over position.} and 
a {\it mutation} operation resulting in a new individual which is a 
member of the next generation. 
Such a breeding is repeated until the next generation has been formed. 
Finally, the whole process of creating new generations is repeated 
iteratively until the population confines one or several regions of 
sufficiently high fitness in parameter space. For more details see 
Theis (1999) or Theis \& Kohle (2001).

   In addition to the advantage of GAs to be almost independent on
the initial (randomly) chosen population and, hence, their ability to leave
local optima, GAs are easily parallelized. Using a master-slave technique
a single fit can be done in a few CPU minutes on a CRAY T3E or a large
Beowulf cluster (Theis \& Harfst 2000).

%
%
\section{Towards an automatic fitting procedure}
\label{automaticfitting}

\subsection{Recovering artificial intensity maps}

   This section describes two tests. In the first test we use a self-consistent
model of NGC 4449. We ''observe'' it with a numerical telescope and use
the derived FITS image as an input for our GA. Contrary to previous
applications this task checks the applicability of FITS images
as an input source of the GA. Additionally (and more important), it 
tests whether the restricted N-body models can be used to fit a 
system with an independently derived map: here differences between the 
restricted N-body and the self-consistent simulations concern e.g.\ 
the treatment 
of the halo and the initial setup of the disk (no dark halo and 
-- initially -- only purely 
circular motions in the case of restricted N-body simulations).

\begin{figure*}[htbp]
  \centerline{
  \psfig{figure=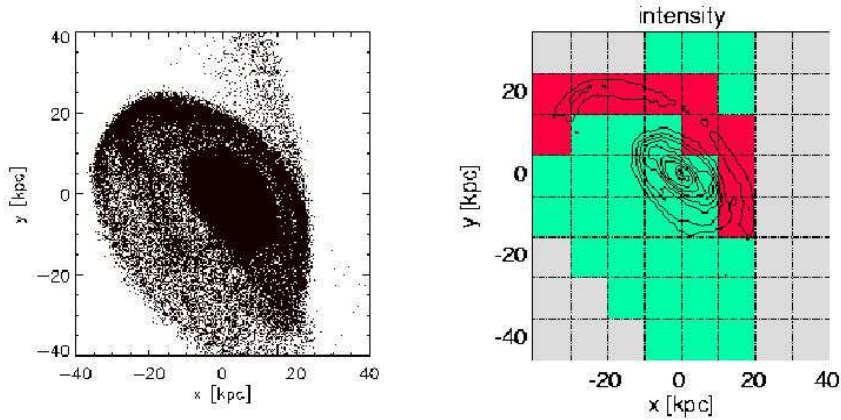,width=12cm,angle=270.0}
   }
  \caption[]{A numerical model for NGC 4449: particle configuration
   (left) and derived intensity map (right). The different grey scales
   on the fitness grid (intensity map) correspond to different weight
   factors used for the GA fitting procedure.}
   \label{fig_4449}
\end{figure*}

\begin{table*}[htbp]
   \begin{center}
   \caption{GA fits derived from an artificial reference map of 
     NGC 4449 using different statistical weight}
   \label{tabweight}
   \vspace*{0.2cm}

   \begin{tabular}{| l || c | c | c | c |}
   {\bf parameter} & reference & equal weight & $W = 4$ & $W = 100$  \\ \hline
       eccentricity          & 0.5   & 0.88  & 0.52  & 0.5           \\
       mass ratio            & 0.2   & 0.11  & 0.17  & 0.13          \\
       perigalactic distance & 25.0  & 18.2  & 23.3  & 24.1          \\
       orbital inclination   & 40.0  & 25.0  & 42.7  & 45.8          \\
       inclination of disc   & 60.0  & 53.9  & 58.2  & 58.5          \\
       P.A. of disc          & 230.0 & 234.8 & 226.1 & 230.2         \\ \hline
   \end{tabular}
   \end{center}
\end{table*}

  The self-consistent
numerical model is identical to the $e=0.5$ model of Theis \& Kohle (2001). 
Fig.\ \ref{fig_4449} shows the particle distribution of the N-body simulation
and the derived intensity map (contour lines). The comparison between
this reference map and the numerical simulations, i.e.\ the
determination of the fitness, is performed by 
$f = 1 / \sum w_i d_i$, where the sum extends over all grid cells.
$d_i$ quantifies the deviation between both maps. It can be calculated
e.g.\ from relative or absolute differences of the intensities.
$w_i$ is the statistical weight of a cell. If all cells have equal weight,
we get here a poor fit, as Tab.\ \ref{tabweight} demonstrates (though some
parameters like the orientation of the disk are nicely reproduced).

A much better fit can be achieved, if the tidally induced structures
get a larger statistical weight. As an example
we split the fitness grid into three regions
(Fig.\ \ref{fig_4449}, right): One region contains the main
tidal feature, the streamers (dark), another one covers the region where 
galactic matter is found ($w_i=1$) and, finally,
there is a region without any particles. Already a weight of $w_i=4$ of the
streamers improves the result substantially: all parameters are
recovered within 15\% accuracy, most of them even much better.

\subsection{Recovering observational intensity maps}

\begin{figure*}[phtb]
  \centerline{
  \psfig{figure=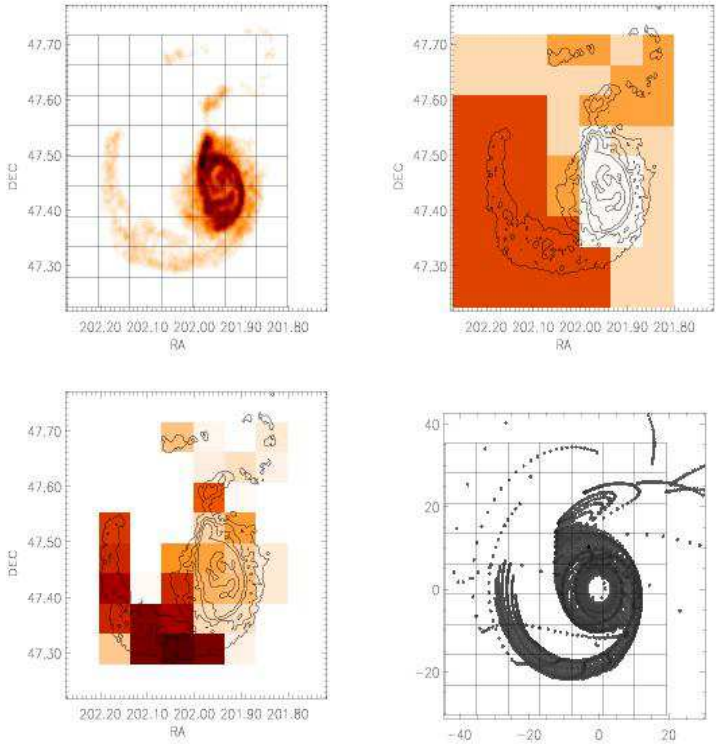,width=12cm,angle=0.0}
  }
  \caption[]{GA analysis of HI data of M 51. Shown are the original HI 
    data superimposed on the fitness grid (upper left), the areas selected
    for different statistical weight (upper right),
    the intensity values on the fitness grid after convolution of intensities
    with the statistical weight (lower left) and a fit found by the genetic
    algorithm after 12 generations (lower right).}
  \label{fig_m51}
\end{figure*}

   As a second test we used HI observations of M 51 
(Rots et al.\ 1990) as a direct input for our genetic algorithm. The HI data
are shown in the upper left diagram of Fig.\ \ref{fig_m51}. The 
centers of both galaxies are in the dark area covering only two adjacent
grid cells. However, the HI distribution is much more extended. 
Assuming that these structures are formed by galactic tides, they 
should be the prime targets for modeling the interaction of 
NGC 5194 and NGC 5195. Therefore, we increased their statistical weight 
in the GA analysis. 

  Typically, the GA easily recovers the extended
tidal arm in the south-east. Additionally, HI is found north of NGC 5195, 
though its detailed structure is more difficult to model. Different
to the artificial maps created by an N-body simulation, the error cannot
be estimated from the known solution. Therefore, we compared the results
of a small series of different GA runs. The relative deviation of the
derived parameters varies between a few percent (the orbital inclination)
up to 25\% (e.g.\ eccentricity, mass ratio of the galaxies). It is remarkable
that all models are characterized by a highly inclined
($i=73^\circ \pm 3^\circ$), elliptical ($e=0.72 \pm 0.15$) orbit.
The mass ratio of both galaxies is $q=0.29 \pm 0.06$.

%
%
\section{Summary}
\label{summary}

    We have demonstrated that GA-based fitting strategies cannot
only be used to check the uniqueness of preferred interaction
scenarios, but also for an automatic fit of given observational 
high-quality data. Different to uniqueness tests
where the same numerical procedures are applied for the generation
of reference maps and the GA fitting procedure, it is -- at least --
useful (if not necessary) to emphasize tidally affected structures by 
a larger statistical weight. In that case the GA is able to reproduce
given intensity maps (here for M 51 and NGC 4449). The interaction parameters
are determined within a statistical error of 25\% or better. 

   As a next step we plan to use the velocity information for the
fitness calculation. At least for some interacting galaxies, we expect
to get better constrained models, as Salo \& Laurikainen (2000) 
demonstrated in the case of M 51.\\

\noindent
{\bf Acknowledgements.}
The authors are grateful to Paul Charbonneau and 
Barry Knapp for providing their (serial) genetic algorithm {\sc pikaia}.
Additionally, we thank Albert Bosma and Lia Athanassoula for 
providing us with the HI data of M 51. The artificial intensity map
of the self-consistent simulation was 
created with the NEMO N-body analysis package.

%
%
{\small
\begin{description}{} \itemsep=0pt \parsep=0pt \parskip=0pt \labelsep=0pt
\item {\bf References}

\item Charbonneau P., 1995, Astrophys.\ J. Supp., 101, 309
\item Holland J., 1975, {\it Adaptation in natural and artificial systems},
         Univ.\ of Michigan Press, Ann Arbor
\item Rots A.H., Bosma A., van der Hulst J.M. et al., 1990, Astron.\ J.,
          100, 387
\item Salo H., Laurikainen E., 2000, Mon.\ Not.\ R. Astron.\ Soc., 319, 377
\item Theis Ch., 1999, Rev.\ Mod.\ Astron., 12, 309
\item Theis Ch., Harfst S., 2000, in {\it Dynamics of Galaxies: from the
    Early Universe to the Present}, F. Combes et al.\ (eds.), ASP conf.\
    ser.\ 197, p.\ 357
\item Theis Ch., Harfst S., 2001, in Proc.\ of {\it Evolution of Galaxies.
  I -- Observational clues}, Granada, J. Vilches et al.\ (eds.), in press
\item Theis Ch., Kohle S., 2001, Astron.\ Astrophys., in press
  (see also astro--ph/0104304)
\item Wahde M., 1998, Astron.\ Astrophys.\ Supp., 132, 417

\end{description}
}

\end{document}